\documentclass[preprint,showpacs,preprintnumbers]{revtex4}
\usepackage{amssymb}
\usepackage{amsmath}
\usepackage{graphicx}
\usepackage{dcolumn}
\usepackage{bm}

\input{tcilatex}
\begin{document}

\title{Robust zero-averaged wave-number gap inside gapped graphene
superlattices}
\author{Li-Gang Wang$^{1,2}$, Xi Chen$^{3,4}$ }
\affiliation{$^{1}$Department of Physics, Zhejiang University, 310027 Hangzhou, China\\
$^{2}$Centre of Optical Sciences and Department of Physics, The Chinese
University of Hong Kong, Shatin, N. T., Hong Kong, China\\
$^{3}$Department of Physics, Shanghai University, 200444 Shanghai, China\\
$^{4}$Departamento de Qu\'{\i}mica-F\'{\i}sica, UPV-EHU, Apdo 644, 48080
Bilbao, Spain}

\begin{abstract}
In this paper, the electronic band structures and its transport properties
in the gapped graphene superlattices, with one-dimensional (1D) periodic
potentials of square barriers, are systematically investigated. It is found
that a zero averaged wave-number (zero-$\overline{k}$ ) gap is formed inside
the gapped graphene-based superlattices, and the condition for obtaining
such a zero-$\overline{k}$ gap is analytically presented. The properties of
this zero-$\overline{k}$ gap including its transmission, conductance and
Fano factor are studied in detail. Finally it is revealed that the
properties of the electronic transmission, conductance and Fano factor near
the zero-$\overline{k}$ gap are very insensitive to the structural disorder
for the finite graphene-based periodic-barrier systems.
\end{abstract}

\maketitle

\section{Introduction}

Graphene, a single layer of carbon atoms densely packed in a honeycomb
lattice, has attracted a lot of research interest due to its remarkable
electronic properties and its potential applications \cite%
{Novoselov2004,Zhang2005,Novoselov2005a,Berger2006,Katsnelson2006,reviews}.
Inside the pristine graphene, the low-energy charge carriers can be formally
described by a massless Dirac equation, and near Dirac point one has
discovered many intriguing properties, such as the unusual energy
dispersion, the chiral behavior \cite{Katsnelson2006,Katsnelson2006b},
ballistic charge transport \cite{Miao2007,Du2008}, Klein tunneling \cite%
{Cheianov2006}, and unusual quantum Hall effect \cite%
{Novoselov2005a,Purewal2006,Neto2006}, bipolar supercurrent \cite%
{Heersche2007}, frequency-dependent conductivity \cite{Kuzmenko2008},
gate-tunable optical transitions \cite{Wang2008}, and so on.

However, for applications of graphene to nanoelectronics, it is crucial to
generate a band gap in Dirac spectrum in order to control the electronic
conductivity, such as a channel material for field-effect transistor. For
realizing this purpose, several approaches are studied both theoretically
and experimentally. One of them is using the quantum confinement effect in
graphene nanoribbons \cite{Son2006a,Son2006b,Han2007} and graphene quantum
dots \cite{Trauzettel2007}. It has been shown that the size of the gap
increases as the nanoribbon width decreases and it also strongly depends on
the detailed structure of the ribbon edges. An alternative method is
spin-orbit coupling, which also leads to generate a small gap due to both
intrinsic spin-orbit interaction or the Rashba interaction \cite%
{Kane2005,Min2006,Yao2007}. Another approach is substrated-induced band gaps
for graphene supported on boron nitride \cite{Giovannetti2007} or SiC \cite%
{Zhou2007,Kim2008} by making the two carbon sublattices (A and B
sublattices) inequivalent; and with this approach, a gap of 260meV is
experimentally demonstrated \cite{Zhou2007}. Therefore the quasiparticles in
the graphene grown on a SiC or boron nitride substrate behave differently
from those in the graphene grown on SiO$_{2}$. The effect of sublattice
symmetry breaking on the induced gap is also systematically investigated 
\cite{Qaiumzadeh2009}. There are also theoretical works to engineer the
tunable bandgap by periodic modulations of the graphene lattice via the
hydrogenation of graphene \cite{Duplock2004}, and a recent experiment
demonstrates that patterned hydrogen adsorption on graphene induces a
bandgap of at least 450meV around the Fermi level \cite{Balog2010}.

Since superlattices are very successful in controlling the electronic
structures of many conventional semiconducting materials (e.g. see Ref. \cite%
{Tsu2005}), the devices of graphene-based superlattices has attracted much
attention. It can be the periodic potential structures generated by
different methods, such as electrostatic potentials \cite%
{Bai2007,Park2008b,Barbier2008,Park2008a,Barbier2009,Tiwari2009} and
magnetic barriers \cite{RamezaniMasir2008}. In gapless graphene-based
superlattices, researchers have found that a one-dimensional (1D)
periodic-potential superlattice possesses some distinct electronic
properties, such as the strong anisotropy for the low-energy charge
carriers' group velocities \cite{Park2008b}, the formation of the extra
Dirac points and new zero energy states \cite{Brey2009,Park2008a}, and the
unusual properties of Landau levels and the quantum Hall effect for these
extra Dirac fermions \cite{Park2009}. From the previous studies \cite%
{Barbier2008,Barbier2010,LGWang2010,Arovas2010,Ho2009}, one has known that
for the gapless graphene superlattices there is no gap opening at the normal
incidence due to the Klein tunneling. Most recently, the new electronic
properties in gapped graphene-based devices are discovered since the Klein
tunneling is suppressed due to the presence of a gap \cite%
{Gomes2008,Soodchomshom2009,Jiang2010,Esmailpour2010,Guinea2010}.

All the above investigations stimulate us to study how the electronic
properties and bandgap structures of the grapped graphene superlattices are
affected due to a gap opening at the Dirac point, and what properties are
derived for the gapped graphene superlattices that are different from the
gapless graphene superlattices. In our previous work \cite{LGWang2010}, we
have found that a new Dirac point is formed at the energy which corresponds
to the zero averaged wave-number inside the gapless graphene-based
superlattices. In this paper, we will continue to investigate the electronic
band structures and their transport properties for the gapped graphene
superlattices with 1D periodic potentials of square barriers. We find that a
zero averaged wave-number (zero-$\overline{k}$) gap is formed inside the
gapped graphene-based superlattices, which is very similar to the
zero-averaged refractive-index gap in 1D photonic crystals consisted of
left-handed and right-handed materials \cite{Li2003}. The properties of this
zero-$\overline{k}$ gap are detailed studied, and the related electronic
transmission, conductance and Fano factor near the zero-$\overline{k}$ gap
in the finite graphene superlattices are further illustrated.

The outline of this paper is the following. In Sec. II, we introduce a
transfer matrix method to calculate the reflection and transmission for the
gapped graphene superlattices. In Sec. III, we first discuss the electronic
band structures for the infinite gapped graphene-based periodic-barrier
superlattices, and then we investigate the changes of the transmission,
conductance and Fano factor for the finite superlattices and the effects of
the structural disorders on the electronic properties are also discussed in
detail. Finally, in Sec. IV, we summarize our results.

\section{Transfer Matrix method for the gapped mono-layer graphene
superlattices}

We consider a mono-layer graphene with a peculiar gap due to the sublattice
symmetry breaking or the intrinsic spin-orbit interaction. In this
situation, the Hamiltonian of an electron in the presence of the
electrostatic potential $V(x)$, which only depends on the coordinate $x$, is
given by \cite{Esmailpour2010,Qaiumzadeh2009b}%
\begin{equation}
\hat{H}=v_{F}\mathbf{\sigma }\cdot \mathbf{p}+V(x)\hat{I}+\Delta \sigma _{z},
\label{Hamiltonian}
\end{equation}%
where $\mathbf{p}=(p_{x},p_{y})=(-i\hbar \frac{\partial }{\partial x}%
,-i\hbar \frac{\partial }{\partial y})$ is the momentum operator with two
components, $\mathbf{\sigma }=(\sigma _{x},\sigma _{y})$, and $\sigma
_{x},\sigma _{y}$ and $\sigma _{z}$ are Pauli matrices, $\hat{I}$ is a $%
2\times 2$ unit matrix, and $v_{F}\approx 10^{6}$m/s is the Fermi velocity.
Here $\Delta =mv_{F}^{2}$ is the energy gap due to the sublattice symmetry
breaking \cite{Zhou2007}, or $\Delta =\Delta _{SO}$ is the energy gap due to
the intrinsic spin-orbit interaction \cite{Kane2005}. From the experimental
data, we know that the maximum energy gap could be $\sim $260meV due to the
sublattice symmetry breaking \cite{Zhou2007}.

The above Hamiltonian acts on the state of a two-component pseudo-spinor, $%
\Psi =(\tilde{\psi}_{A},\tilde{\psi}_{B})^{T},$ where $\tilde{\psi}_{A}$ and 
$\tilde{\psi}_{B}$ are the smooth enveloping functions for two triangular
sublattices in the mono-layer graphene, and the symbol "$T$" denotes the
transpose operator. In the $y$ direction, because of the translation
invariance, the wave functions $\tilde{\psi}_{A,B}(x,y)$ can be factorized
by $\tilde{\psi}_{A,B}(x,y)=\psi _{A,B}(x)e^{ik_{y}y}.$ Therefore, from Eq. (%
\ref{Hamiltonian}), we obtain%
\begin{eqnarray}
\frac{d\psi _{A}}{dx}-k_{y}\psi _{A} &=&i\eta _{+}\psi _{B},  \label{SD1} \\
\frac{d\psi _{B}}{dx}+k_{y}\psi _{B} &=&i\eta _{-}\psi _{A},  \label{SD2}
\end{eqnarray}%
where $\eta _{\pm }=[E-V(x)\pm \Delta ]/(\hbar v_{F})$ are the transit (or
coupled) parameters from $\psi _{B}$ ($\psi _{A}$) to $\psi _{A}$ ($\psi
_{B} $), $E$ is the incident electron energy, and $k_{0}=E/\hbar v_{F}$
corresponds to the incident electronic wavenumber. When $\Delta \rightarrow
0 $, the above two equations reduce to the cases in Refs. \cite%
{Barbier2010,LGWang2010,Arovas2010}.

For the gapped graphene superlattices, we assume that the potential $V(x)$
is comprised of periodic potentials of square barriers as shown in Fig. 1.
Inside the $j$ th barrier, $V_{j}(x)$ is a constant, therefore, from Eqs. (%
\ref{SD1}) and (\ref{SD2}), we have%
\begin{eqnarray}
\frac{d^{2}\psi _{A}}{dx^{2}}+(k_{j}^{2}-k_{y}^{2})\psi _{A} &=&0,
\label{HoM1} \\
\frac{d^{2}\psi _{B}}{dx^{2}}+(k_{j}^{2}-k_{y}^{2})\psi _{B} &=&0,
\label{HoM2}
\end{eqnarray}%
where $k_{j}=$sign$(\eta _{j+})[(E-V_{j})^{2}-\Delta ^{2}]^{1/2}/(\hbar
v_{F})$ is the wavevector inside the barrier $V_{j}$ for the case of $%
|E-V_{j}|>\Delta $, otherwise $k_{j}=i[\Delta
^{2}-(E-V_{j})^{2}]^{1/2}/(\hbar v_{F})$; meanwhile we always have the
relation $\eta _{j+}\cdot \eta _{j-}=k_{j}^{2}$. Here the subscript "$j$"
denotes the quantities inside the $j$ th barrier, and $j=0,1,2,3,\cdots
,2N,e $, where $j=0$ denotes the incident region, $j=e$ denotes the exit
region, and $N$ is the periodic number. Note that $k_{j}$ is negative in the
case of $\eta _{j+}<0$, which leads to the electron's "Veselago Lens" \cite%
{Cheianov2007}.

Following the calculation method in Ref. \cite{LGWang2010}, we can readily
obtain the relation between $\binom{\psi _{A}(x_{j-1})}{\psi _{B}(x_{j-1})}$
and $\binom{\psi _{A}(x_{j-1}+\Delta x)}{\psi _{B}(x_{j-1}+\Delta x)}$ in
the following form:%
\begin{equation}
\binom{\psi _{A}(x_{j-1}+\Delta x)}{\psi _{B}(x_{j-1}+\Delta x)}%
=M_{j}(\Delta x,E,k_{y})\binom{\psi _{A}(x_{j-1})}{\psi _{B}(x_{j-1})},
\end{equation}%
where the transfer matrix $M_{j}$ is given by%
\begin{equation}
M_{j}(\Delta x,E,k_{y})=\left( 
\begin{array}{cc}
\frac{\cos (q_{j}\Delta x-\theta _{j})}{\cos \theta _{j}} & i\frac{\sin
(q_{j}\Delta x)}{p_{j}\cos \theta _{j}} \\ 
i\frac{p_{j}\sin (q_{j}\Delta x)}{\cos \theta _{j}} & \frac{\cos
(q_{j}\Delta x+\theta _{j})}{\cos \theta _{j}}%
\end{array}%
\right) ,  \label{MMT}
\end{equation}%
which denotes the characteristic matrix for the two-component wave function
propagating from the position $x_{j-1}$ to another position $x_{j-1}+\Delta
x $ inside the $j$ th barrier. Here $p_{j}=\eta _{j-}/k_{j}$, $q_{j}=$sign$%
(\eta _{j+})\sqrt{k_{j}^{2}-k_{y}^{2}}$ is the $x$ component of the
wavevector inside the $j$ th barrier for $k_{j}^{2}>k_{y}^{2}$, otherwise $%
q_{j}=i\sqrt{k_{y}^{2}-k_{j}^{2}}$, and $\theta _{j}=$arcsin($k_{y}/k_{j}$)
is the angle between two components $q_{j}$ and $k_{y}$ inside the $j$ th
barrier. When $\Delta =0$, we have $\eta _{j+}=\eta _{j-}=k_{j}$, so that $%
p_{j}=1$ for the gapless mono-layer graphene ($\Delta =0$), which leads the
transfer matrix (\ref{MMT}) to be the same as that in Ref. \cite{LGWang2010}%
. Here we would like to point out that, in the case of $\eta _{j+}=0$, the
transfer matrix (\ref{MMT}) should be replaced by 
\begin{equation}
M_{j}(\Delta x,E,k_{y})=\left( 
\begin{array}{cc}
\exp (k_{y}\Delta x) & 0 \\ 
ip_{j}\sinh (k_{y}\Delta x) & \exp (-k_{y}\Delta x)%
\end{array}%
\right) ,  \label{MMT1}
\end{equation}%
and in this case, $p_{j}=\eta _{j-}/k_{y}$. When $\eta _{j-}=0$, the
transfer matrix (\ref{MMT}) should be 
\begin{equation}
M_{j}(\Delta x,E,k_{y})=\left( 
\begin{array}{cc}
\exp (k_{y}\Delta x) & ip_{j}\sinh (k_{y}\Delta x) \\ 
0 & \exp (-k_{y}\Delta x)%
\end{array}%
\right) ,  \label{MMT2}
\end{equation}%
where $p_{j}=\eta _{j+}/k_{y}$.

With the knowledge of the transfer matrices (\ref{MMT}, \ref{MMT1}, and \ref%
{MMT2}), we can easily connect the input and output wave functions by the
following equation:

\begin{equation}
\left( 
\begin{array}{c}
\psi _{A}(x_{e}) \\ 
\psi _{B}(x_{e})%
\end{array}%
\right) =\mathbf{X}\left( 
\begin{array}{c}
\psi _{A}(0) \\ 
\psi _{B}(0)%
\end{array}%
\right) ,  \label{WAVE00dd}
\end{equation}%
with the matrix%
\begin{equation}
\mathbf{X}\mathbf{=}\left( 
\begin{array}{cc}
x_{11} & x_{12} \\ 
x_{21} & x_{22}%
\end{array}%
\right) =\dprod\limits_{j=1}^{2N}M_{j}(w_{j},E,k_{y}),  \label{XXmatrix}
\end{equation}%
which is the total transfer matrix of the electron's transport from the
incident\ end ($x=0$) to the exit end ($x=x_{e}$) in the $x$ direction,
where $w_{j}$ is the width of the $j$ th potential barrier.

For obtaining the transmission and reflection coefficients, we should build
up the boundary condition. As shown in Fig. 1, we assume that a free
electron of energy $E$ is incident from the region $x\leq 0$ at any incident
angle $\theta _{0}$. In this region, the electronic wave function is a
superposition of the incident and reflective wave packets, so at the
incident end ($x=0$) we have%
\begin{equation}
\left( 
\begin{array}{c}
\psi _{A}(0) \\ 
\psi _{B}(0)%
\end{array}%
\right) =\left( 
\begin{array}{c}
1+r \\ 
p_{0}(e^{i\theta _{0}}-re^{-i\theta _{0}})%
\end{array}%
\right) \psi _{i}(E,k_{y}),  \label{WAVE00}
\end{equation}%
where $r$ is the reflection coefficient, $p_{0}$ is the quantity in the
incident region, and $\psi _{i}(E,k_{y})$ is the incident wavepacket of the
electron at $x=0$.

At the exit end ($x=x_{e}$), we have%
\begin{equation}
\left( 
\begin{array}{c}
\psi _{A}(x_{e}) \\ 
\psi _{B}(x_{e})%
\end{array}%
\right) =\left( 
\begin{array}{c}
t \\ 
tp_{e}e^{i\theta _{e}}%
\end{array}%
\right) \psi _{i}(E,k_{y}),  \label{WAVEdd}
\end{equation}%
with the assumption of $\psi _{A}(x_{e})=t\psi _{i}(E,k_{y})$, where $t$ is
the transmission coefficient of the electronic wave function through the
whole structure, $p_{e}$ is the quantity in the exit region, and $\theta
_{e} $ is the exit angle at the exit end. By substituting Eqs. (\ref{WAVE00}%
, \ref{WAVEdd}) into Eq. (\ref{WAVE00dd}), we have the following equations%
\begin{eqnarray}
t &=&(1+r)x_{11}+p_{0}(e^{i\theta _{0}}-re^{-i\theta _{0}})x_{12},
\label{rt1} \\
tp_{e}e^{i\theta _{e}} &=&(1+r)x_{21}+p_{0}(e^{i\theta _{0}}-re^{-i\theta
_{0}})x_{22}.  \label{rt2}
\end{eqnarray}%
After solving the above two equations, we find the reflection and
transmission coefficients%
\begin{eqnarray}
r(E,k_{y}) &=&\frac{(x_{22}p_{0}e^{i\theta _{0}}-x_{11}p_{e}e^{i\theta
_{e}})-x_{12}p_{0}p_{e}e^{i(\theta _{e}+\theta _{0})}+x_{21}}{%
(x_{22}p_{0}e^{-i\theta _{0}}+x_{11}p_{e}e^{i\theta
_{e}})-x_{12}p_{0}p_{e}e^{i(\theta _{e}-\theta _{0})}-x_{21}},
\label{rrcoeff} \\
t(E,k_{y}) &=&\frac{2p_{0}\cos \theta _{0}}{(x_{22}p_{0}e^{-i\theta
_{0}}+x_{11}p_{e}e^{i\theta _{e}})-x_{12}p_{0}p_{e}e^{i(\theta _{e}-\theta
_{0})}-x_{21}},  \label{ttcoeff}
\end{eqnarray}%
where we have used the property of $\det [\mathbf{X}]=1$.

Since the reflection and transmission coefficients are obtained, the total
conductance can also be calculated. Using the B\"{u}ttiker formula,\cite%
{Datta1995} the total conductance of the system at zero temperature is given
by%
\begin{equation}
G=G_{0}\int_{0}^{\pi /2}T(E,k_{y})\cos \theta _{0}d\theta _{0},
\label{COND1}
\end{equation}%
where $T(E,k_{y})=\left\vert t(E,k_{y})\right\vert ^{2}$ is the
transmitivity, $G_{0}=2e^{2}mv_{F}L_{y}/\hbar ^{2}$, and $L_{y}$ is the
width of the graphene strip along the $y$ axis. Furthermore, we can also
study the Fano factor (F) in the gapped graphene superlattices, which is
given by \cite{Tworzydlo2006} 
\begin{equation}
F=\frac{\int_{-\pi /2}^{\pi /2}T(1-T)\cos \theta _{0}d\theta _{0}}{%
\int_{-\pi /2}^{\pi /2}T\cos \theta _{0}d\theta _{0}}.  \label{Fano1}
\end{equation}%
Combining Eqs. (\ref{rrcoeff})-(\ref{Fano1}), the reflection, transmission,
conductance, and Fano factor for the gapped graphene superlattices can be
obtained by the numerical calculations. In the following discussions, we
will discuss the properties of the electronic band structure, transmission,
conductance and Fano factor for the gapped graphene-based periodic
potentials of square barriers.

\section{Results and Discussions}

In this section, first we will discuss the electronic band structures for
the infinite periodic-barrier systems, and then we will discuss properties
of the electronic transmission, conductance and Fano factor for the finite
periodic-barrier systems with or without the structural disorder.

\subsection{Infinite periodic-barrier systems}

First, let us investigate the electronic bandgap structure for an infinite
gapped graphene-based periodic-barrier systems, i.e., ($AB$)$^{N}$, where
the symbols $A$ and $B$ from now on denote the different barriers $A$ and $B$
with the electrostatic potentials $V_{A}$ and $V_{B}$, and the widths $w_{A}$
and $w_{B}$, respectively, and $N$ is the periodic number. We assume $%
V_{A}>V_{B}$. By using the Bloch's theorem, the electronic band structure
for an infinite periodic structures, i.e., ($AB$)$^{N}$ with $N\rightarrow
\infty $, is governed by the following relation:%
\begin{eqnarray}
\cos [\beta _{x}\Lambda ] &=&\frac{1}{2}Tr[M_{A}M_{B}]=\frac{1}{2\cos \theta
_{A}\cos \theta _{B}}\left[ \cos (q_{A}w_{A}-\theta _{A})\cos
(q_{B}w_{B}-\theta _{B})\right.  \notag \\
&&\left. +\cos (q_{A}w_{A}+\theta _{A})\cos (q_{B}w_{B}+\theta _{A})-\left( 
\frac{p_{B}}{p_{A}}+\frac{p_{A}}{p_{B}}\right) \sin (q_{A}w_{A})\sin
(q_{B}w_{B})\right]  \notag \\
&=&\cos (q_{A}w_{A})\cos (q_{B}w_{B})-\frac{\frac{p_{B}}{p_{A}}+\frac{p_{A}}{%
p_{B}}-2\sin \theta _{A}\sin \theta _{B}}{2\cos \theta _{A}\cos \theta _{B}}%
\sin (q_{A}w_{A})\sin (q_{B}w_{B})  \notag \\
&=&\cos [q_{A}w_{A}+q_{B}w_{B}]-\frac{\frac{p_{B}}{p_{A}}+\frac{p_{A}}{p_{B}}%
-2\cos (\theta _{A}-\theta _{B})}{2\cos \theta _{A}\cos \theta _{B}}\sin
(q_{A}w_{A})\sin (q_{B}w_{B}).  \label{DISPERSION11}
\end{eqnarray}%
Here $\Lambda =w_{A}+w_{B}$ is the length of the unit cell.\ Now we assume
that the incident energy of the electron is $V_{B}+\Delta <E<V_{A}+\Delta $,
then we always have $p_{A,B}>0$, $-\pi /2<\theta _{A}<0$, $q_{A}<0$, $%
0<\theta _{B}<\pi /2$, and $q_{B}>0$ for the propagating modes. When $%
-q_{A}w_{A}=q_{B}w_{B}$, the above equation (\ref{DISPERSION11}) becomes%
\begin{equation}
\cos [\beta _{x}\Lambda ]=1+\frac{[\frac{p_{B}}{p_{A}}+\frac{p_{A}}{p_{B}}%
-2\cos (\theta _{A}-\theta _{B})]}{2\cos \theta _{A}\cos \theta _{B}}|\sin
(q_{A}w_{A})|^{2}.  \label{Dis2}
\end{equation}%
Because $\frac{p_{B}}{p_{A}}+\frac{p_{A}}{p_{B}}>2$ (due to $p_{A}\neq
p_{B}\neq 1$ for the gapped graphene superlattices), $\cos (\theta
_{A}-\theta _{B})\leqslant 1$, and $\cos \theta _{A,B}>0$, from the above
equation, we can find that there is no real solution for $\beta _{x}$ when $%
-q_{A}w_{A}=q_{B}w_{B}\neq m\pi $. That is to say, there opens a new band
gap in the gapped graphene-based periodic-barrier structures. At normal
incidence ($\theta _{A}=\theta _{B}=0$), the condition of $%
-q_{A}w_{A}=q_{B}w_{B}\neq m\pi $ (within the energy interval $V_{B}+\Delta
<E<V_{A}+\Delta $) becomes%
\begin{eqnarray}
-k_{A}w_{A} &=&k_{B}w_{B}\neq m\pi ,\text{ \ }  \label{Cond1} \\
\text{or }\ [(E-V_{A})^{2}-\Delta ^{2}]^{1/2}w_{A} &=&[(E-V_{B})^{2}-\Delta
^{2}]^{1/2}w_{B}\neq m\pi .  \label{Cond11}
\end{eqnarray}%
This condition, Eq. (\ref{Cond1}) or (\ref{Cond11}), actually corresponds to
the zero averaged wave number, i.e., $\bar{k}=(k_{A}w_{A}+k_{B}w_{B})/%
\Lambda =0$. Therefore the gap occurring at the zero averaged wave number is
called the zero-averaged wave-number (zero-$\bar{k}$) gap. The distinct
difference between the gapless and gapped graphene superlattices is that for
the gapless case ($\Delta =0$) this zero-$\bar{k}$ gap is close at the
normal incidence sine $p_{A}=p_{B}=1$, while for the gapped case ($\Delta
\neq 0$) it is open even at normal incidence from Eq. (\ref{Dis2}) because
of $\frac{p_{B}}{p_{A}}+\frac{p_{A}}{p_{B}}>2$ (due to $p_{A}\neq p_{B}\neq
1 $). For a special case with equal barrier and well widths, i.e., the ratio 
$w_{A}/w_{B}=1$, from Eq. (\ref{Cond1} or \ref{Cond11}), we can know that
the location of the zero-$\bar{k}$ gap is exactly at $E=(V_{A}+V_{B})/2$.

However, when%
\begin{equation}
-q_{A}w_{A}=q_{B}w_{B}=m\pi  \label{Cond2}
\end{equation}%
is satisfied, then $\sin (q_{A}w_{A})=\sin (q_{B}w_{B})=0$, therefore $\cos
[\beta _{x}\Lambda ]=1$, which tells us that the zero-$\bar{k}$ gap will
begin to be close in the case of normal incidence and a pair of new zero-$%
\bar{k}$ states emerges from $k_{y}=0$ (i.e., the case of inclined
incidence). Actually the above condition (\ref{Cond2}) is the same as that
in 1D photonic crystals consisted of left-handed and right-hand materials 
\cite{LGWang2010b}.

Figures 2(a) to 2(d) show clearly the dependence of the electronic band
structures on the lattice constant $\Lambda $ for the gapped graphene
superlattices with equal barrier and well widths (i.e., the ratio $%
w_{A}/w_{B}=1$). Here we take the parameter $\Delta =5$meV. It is clear seen
that the center of a band gap is exactly at energy $E=40$meV, where the
condition, $q_{A}w_{A}=-q_{B}w_{B}\neq m\pi $, is satisfied, see Figs. 2(a)
and 2(b). The location of this zero-$\bar{k}$ gap is independent of the
lattice constant [see Figs. 2(a) and 2(b)]; while other upper or lower band
gaps are strongly dependent on the lattice constant, and they are shifted
with the changing of the lattice constant. The width of the zero-averaged
wave-number gap depends on the lattice constant, therefore it can be tunable
by changing the lattice constant. For example, in Fig. 2(a), this zero-$\bar{%
k}$ gap has the smallest width of $\sim 7.6$meV, which is larger than the
value of $\Delta $ but smaller than $2\Delta $; and in Fig. 2(b) it has the
smallest width of $\sim 2.6$meV, which is smaller than the value of $\Delta $%
. With the increasing of the lattice constant, the slopes for both the band
edges of the zero-$\bar{k}$ gap become smaller and smaller [see Fig.
2(a-c)], and furthermore the gap is open or close with the change of the
lattice constant [see Figs. 2(e-f)]. In the case when the condition (\ref%
{Cond2}) is valid, the gap is close for the normal incidence ($k_{y}=0$)
[see Fig. 2(c)] or it is close for the inclined incidence ($\pm k_{y}\neq 0$%
) and a pair of two zero-$\bar{k}$ states appear [see Fig. 2(d)]. Actually,
such kind of the crossed points is termed as the extra Dirac points in the
gapless graphene superlattices \cite{Barbier2010,LGWang2010,Arovas2010}.
Compared with Fig. 2(e) and 2(f), it is found that for the inclined cases,
the zero-averaged wave-number gap is enlarged and the extra Dirac points are
occurring at the same energy with those of the touching points in the normal
case.

Similarly, figure 3 shows the change of the electronic band structure for
the gapped graphene superlattices with unequal barrier and well widths
(i.e., the ratio $w_{A}/w_{B}\neq 1$). From Figs. 3(a) to 3(c), it is clear
that the position of zero-$\bar{k}$ gap is still independent of the lattice
constant, and in this example it is located at $E\approx 45.62$meV, i. e.,
the one of roots for the condition (\ref{Cond1}) or (\ref{Cond11}). By the
way, another root of the condition (\ref{Cond1}) or (\ref{Cond11}) is
unphysical since it is outside of the interval $V_{B}+\Delta <E<V_{A}+\Delta 
$. Meanwhile, the width of this zero-$\bar{k}$ gap could be still adjusted
by changing its lattice constant with the fixed ratio $w_{A}/w_{B}$. However
for the other upper and lower band gaps are strongly shifted due to the
change of the lattice constant. In Fig. 3(c), it is also occurring the
touching effect of the upper and lower bands due to that the condition (\ref%
{Cond2}) is satisfied at $k_{y}=0$. Different from the above case with equal
barrier and well widths ($w_{A}/w_{B}=1$), in Fig. 3(d), when the lattice
constant increases larger, the touch points move down for the inclined
incidence ($\pm k_{y}\neq 0$) in the cases with $w_{A}/w_{B}>1$. In the
cases with $w_{A}/w_{B}<1$, one can find that the touching points move up
for the inclined incidence [see Fig. 3(e)]. This property of the touch
points, depending on the ratio of widths\ $w_{A}/w_{B}$, is similar to that
in the gapless graphene superlattices \cite{Barbier2010}. It should be
pointed out that the touch effects in Figs. 2(c) and 3(c) is very similar to
the cases of zero-width band gap associated with the zero-averaged
refractive index in photonic crystals containing left-handed materials \cite%
{LGWang2010b,DiosLeyva2009}.

\subsection{Finite periodic-barrier systems}

Now let us turn to discuss the properties of the transmission, conductance
and Fano factor in the finite periodic-barrier systems. In order to know the
information of the band structures for the finite systems, we have to
calculate the transmission as functions of the incident electron energy and
the incident angle. Figure 4 shows the transmission properties of an
electron passing through $(AB)^{20}$ under different values of $\Delta $.
Figures 5(a) and 5(b) show the changes of electronic conductance and Fano
factor for those cases corresponding to different situations in Fig. 4. It
is clear seen that when $\Delta =0$, the electronic transmission at normal
incidence is always equal to unit, see Fig. 4(a). This property is a
reflection of "Klein tunneling" in the systems of gapless graphene
superlattices \cite%
{Barbier2008,Barbier2009,Barbier2010,LGWang2010,Arovas2010}. From Fig. 4(a),
one can also find that there is a band gap opening up at all inclined angles
around the energy of $E=40$meV. Actually this gap is already termed as the
zero-$\bar{k}$ gap, which is associated with the new Dirac point, in our
previous study on the gapless graphene superlattices \cite{LGWang2010}. In
Fig. 5(a) and 5(b), the conductance is largest and the Fano factor is equal
to 1/3 at the new Dirac point ($E=40$meV) for the case of $\Delta =0$, which
recovers the result for the diffusive behavior near the new Dirac point.
With the increasing of $\Delta $, this band gap gradually opens up at the
normal incidence, see Figs. 4(b)-4(d). Therefore the conductance becomes
smaller and smaller when the gap is open at the normal incidence, and the
Fano factor is enhanced to be larger than 1/3. From Fig. 5(a) and 5(b), one
can see that when the gap is completely open for the larger value of $\Delta 
$, the Fano factor is close to unit because there is no allowed states for
the electrons in this zero-$\bar{k}$ gap. Another remarkable property is
that near the edges of the zero-$\bar{k}$ gap the Fano factor is also larger
than 1/3 for the larger value of $\Delta $, which does not happen in the
cases for smaller values of $\Delta $. It means that for the gapped graphene
superlattices the electron's transport has a distinct difference from the
cases of gapless graphene superlattices. It should also be pointed out that
the differences for the conductances of the higher band gaps in Fig. 5(a)
are much small due to that the passing band is highly shifted to the higher
energy for the inclined incident angles [see Figs 4(a) to 4(d)], and the
Fano factor for the higher band gap in Fig. 5(b) is larger than 1/3 even for
the case of $\Delta =0$.

At last, we consider the effect of the structural disorder on the
transmission of an electron passing through a finite gapped graphene-based
periodic-barrier structure with the width deviation. Here we consider the
periodic-barrier structure $(AB)^{30}$, as an example, with $\Delta =5$meV
and $w_{A}=w_{B}=(20+R)$nm, where $R$ is a random number. Figure 6 shows the
effect of the structural disorder on electronic transmitivities,
conductances and Fano factors. It is clear seen that the zero-$\bar{k}$ gap
is insensitive to the structural disorder, while the other band gaps are
destroyed by strong disorder, see Fig. 6(a). The robustness of the zero-$%
\bar{k}$ gap comes from the fact that the zero-$\bar{k}$ solution remains
invariant under disorder that is symmetric ($+$ and $-$ equally probable),
see Eq. (\ref{DISPERSION11}). These results are very similar to that cases
in the gapless graphene superlattices \cite{LGWang2010}, and the unique
difference is that at the normal incidence the zero-$\bar{k}$ gap is close
for the gapless case while it is still open for the gapped case. From Figs.
6(b) and 6(c), one can find that the structural disorder does strongly
affect on the properties of the electronic conductance and Fano factor when
the incident electron's energy $E$ is far away from the zero-$\bar{k}$ gap.
In Figs. 5(b,c), it is clear that the curves of the conductances and Fano
factors are much different from each other for those energies far away from $%
E=40$meV. Therefore, the zero-$\bar{k}$ gap and its related properties are
very insensitive to the structural disorder while the other bands and gaps
are highly affected by the structural disorder.

\section{Conclusions}

In summary, we have studied the electronic band structures and its transport
properties for the gapped graphene superlattices consisted of 1D periodic
potentials of square barriers. We have found that there is a zero-$\overline{%
k}$ gap inside the gapped graphene-based superlattices, and the location of
the zero-$\overline{k}$ gap is independent of the lattice constant but
depends on the ratio of the barrier and well widths. Furthermore we have
shown that the width of the zero-$\overline{k}$ gap could be controllable by
changing the lattice constant of the gapped superlattices, and under the
certain condition the zero-$\overline{k}$ gap could be close at the case of
normal incidence and the band-crossing phenomena (the extra new Dirac
points) occurs at the case of inclined incidence inside the gapped graphene
superlattices, which are similar to the cases of the gapless graphene
superlattices \cite{Barbier2010,LGWang2010}. Finally it is revealed that the
properties of the electronic transmission, conductance and Fano factor near
the zero-$\overline{k}$ gap are insensitive to the structural disorder
inside the finite periodic-barrier systems. Our analytical and numerical
results on the electronic band structures and their related properties of
the gapped graphene-based superlattices are hopefully of benefit to more
potential applications of graphene-based devices.

\begin{acknowledgments}
This work is supported by the National Natural Science Foundation of China
(10604047 and 60806041), Hong Kong RGC Grant No. 403609 and CUHK 2060360,
the Shanghai Rising-Star Program (No. 08QA14030), the Science and Technology
Commission of Shanghai Municipal (No. 08JC14097), and the Shanghai Leading
Academic Discipline Program (No. S30105). X. C. also acknowledges Juan de la
Cierva Programme, IT 472-10, and FIS2009-12773-C02-01.
\end{acknowledgments}

\newpage

\begin{center}
{\Huge Figures Captions:}
\end{center}

Fig. 1. (Color online). (a) Schematic of a gapped graphene superlattice with
periodic electrodes. (b) Schematic diagram of the electronic spectrum of the
gapped graphene superlattice, and the pink dotted line denotes the periodic
potentials of squared barriers.

Fig. 2. (Color online). Electronic band structures for the gapped graphene
superlattices with equal barrier and well widths ($w_{A}/w_{B}=1$): (a) $%
\Lambda =40$nm, (b) $\Lambda =80$nm, (c) $\Lambda =104.214$nm, and (d) $%
\Lambda =120$nm; and dependence of the band-gap structure on the lattice
constant $\Lambda $ with a fixed transversal wave number: (e) $k_{y}=0$ and
(f) $k_{y}=0.015$nm$^{-1}$. The other parameters are $\Delta =5$meV, $%
V_{A}=80$meV and $V_{B}=0$.

Fig. 3. (Color online). Electronic band structures for the gapped graphene
superlattices with unequal barrier and well widths ($w_{A}/w_{B}=4/3$): (a) $%
\Lambda =35$nm, (b) $\Lambda =70$nm, (c) $\Lambda =106.4$nm, and (d) $%
\Lambda =119$nm; and (e) electronic band structures for the case of $%
w_{A}/w_{B}=3/4$ and $\Lambda =119$nm. The other parameters are the same as
in Fig. 2.

Fig. 4. (Color online). Effects of the parameter $\Delta $ on the electronic
transmission for the finite structure $(AB)^{20}$, (a) $\Delta =0$, (b) $%
\Delta =1$meV, (c) $\Delta =2$meV, and (d) $\Delta =5$meV. The other
parameters are $\Lambda =40$nm, $w_{A}/w_{B}=1$, $V_{A}=80$meV, and $V_{B}=0$%
.

Fig. 5. (Color online). The effects of the parameter $\Delta $ on (a)
conductance and (b) Fano factor. The other parameters are the same as in
Fig. 4.

Fig. 6. (Color online). The effect of the structural disorder on (a)
electronic transimitivity $T=|t|^{2}$, (b) conductance $G/G_{0}$, and (c)
Fano factor, for the gapped graphene superlattice with $\Delta =5$meV, $%
V_{A}=80$meV and $V_{B}=0,$ and $w_{A}=w_{B}=(20+R)$nm, where $R$ is a
random number. The short-dashed lines denote for the structure without
disorder, the dashed lines for $R\in (-2.5,2.5)$nm, the dashed-dot lines for 
$R\in (-5,5)$nm, and the solid lines for $R\in (-7.5,7.5)$nm.

\newpage

\begin{figure}[b]
\centering
\includegraphics[width=10cm]{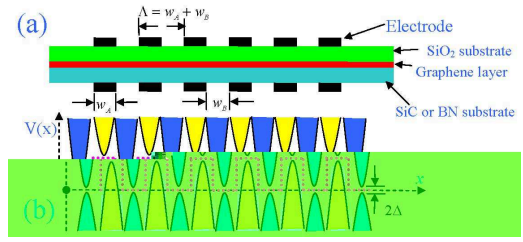}
\caption{ (Color online). (a) Schematic of a gapped graphene superlattice
with periodic electrodes. (b) Schematic diagram of the electronic spectrum
of the gapped graphene superlattice, and the pink dotted line denotes the
periodic potentials of squared barriers.}
\label{fig:FIG1}
\end{figure}

\begin{figure}[b]
\centering
\includegraphics[width=12cm]{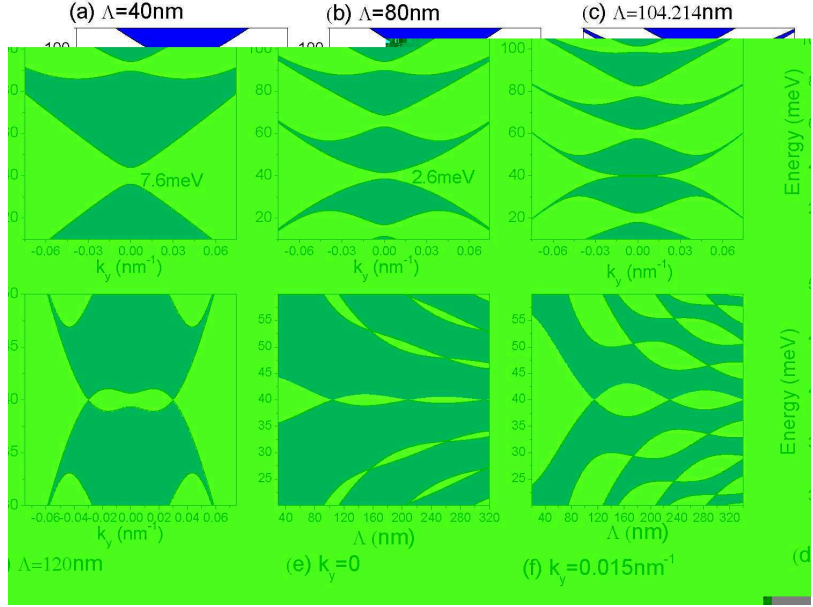}
\caption{(Color online). Electronic band structures for the gapped graphene
superlattices with equal barrier and well widths ($w_{A}/w_{B}=1$): (a) $%
\Lambda =40$nm, (b) $\Lambda =80$nm, (c) $\Lambda =104.214$nm, and (d) $%
\Lambda =120$nm; and dependence of the band-gap structure on the lattice
constant $\Lambda $ with a fixed transversal wave number: (e) $k_{y}=0$ and
(f) $k_{y}=0.015$nm$^{-1}$. The other parameters are $\Delta =5$meV, $%
V_{A}=80$meV and $V_{B}=0$.}
\label{fig:FIG2}
\end{figure}

\begin{figure}[b]
\centering
\includegraphics[width=12cm]{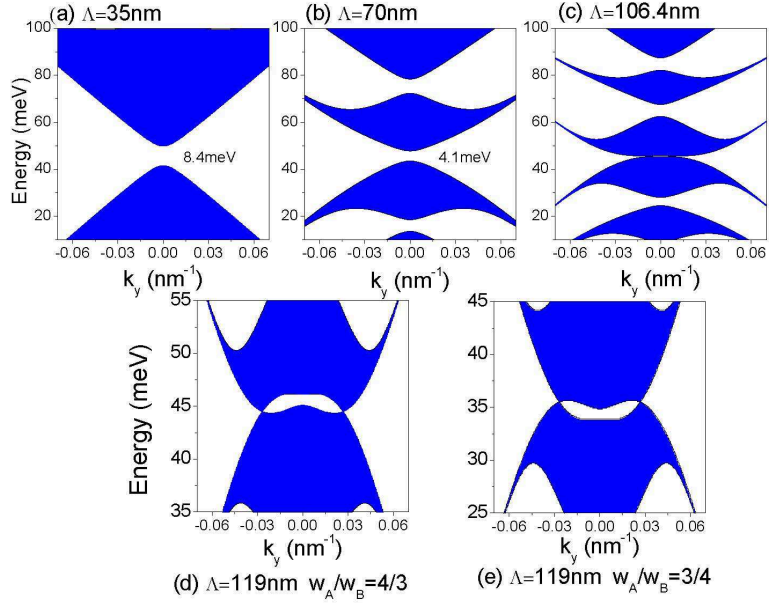}
\caption{(Color online). Electronic band structures for the gapped graphene
superlattices with unequal barrier and well widths ($w_{A}/w_{B}=4/3$): (a) $%
\Lambda =35$nm, (b) $\Lambda =70$nm, (c) $\Lambda =106.4$nm, and (d) $%
\Lambda =119$nm; and (e) electronic band structures for the case of $%
w_{A}/w_{B}=3/4$ and $\Lambda =119$nm. The other parameters are the same as
in Fig. 2.}
\label{fig:FIG3}
\end{figure}

\begin{figure}[b]
\centering
\includegraphics[width=10cm]{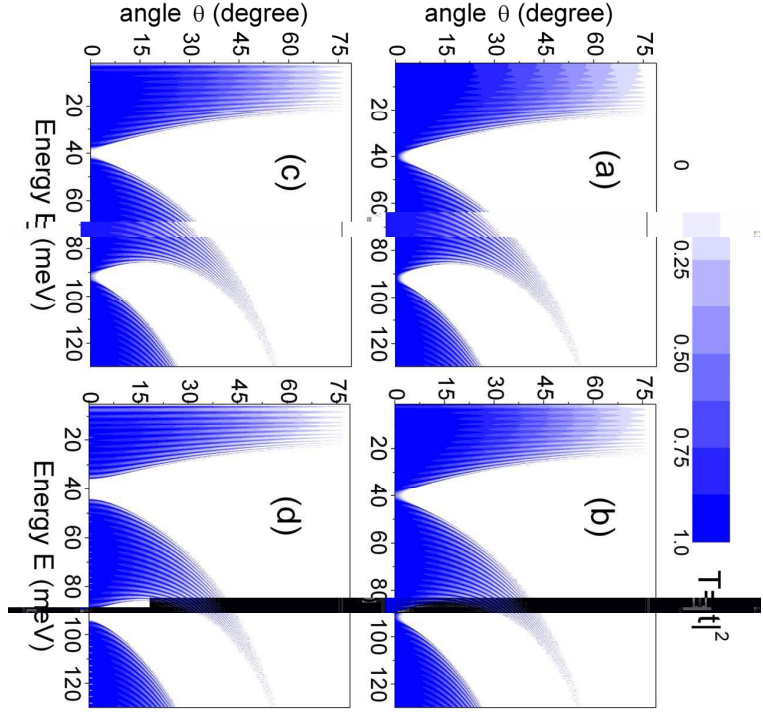}
\caption{(Color online). Effects of the parameter $\Delta $ on the
electronic transmission for the finite structure $(AB)^{20}$, (a) $\Delta =0$%
, (b) $\Delta =1$meV, (c) $\Delta =2$meV, and (d) $\Delta =5$meV. The other
parameters are $\Lambda =40$nm, $w_{A}/w_{B}=1$, $V_{A}=80$meV, and $V_{B}=0$%
.}
\label{fig:FIG4}
\end{figure}

\begin{figure}[b]
\centering
\includegraphics[width=10cm,height=13cm]{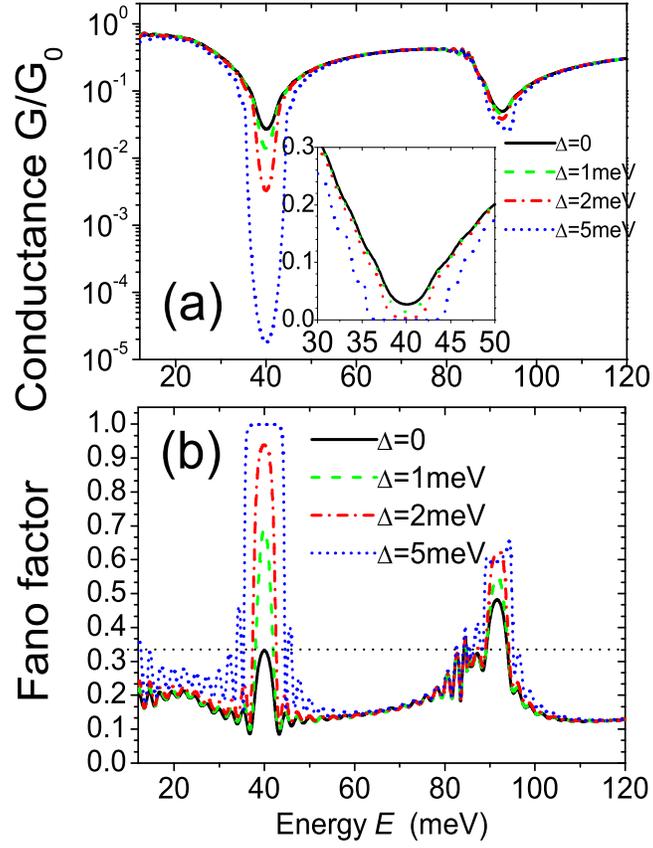}
\caption{(Color online). The effects of the parameter $\Delta $ on (a)
conductance and (b) Fano factor. The other parameters are the same as in
Fig. 4. }
\label{fig:FIG5}
\end{figure}

\begin{figure}[b]
\centering
\includegraphics[width=10cm,height=16cm]{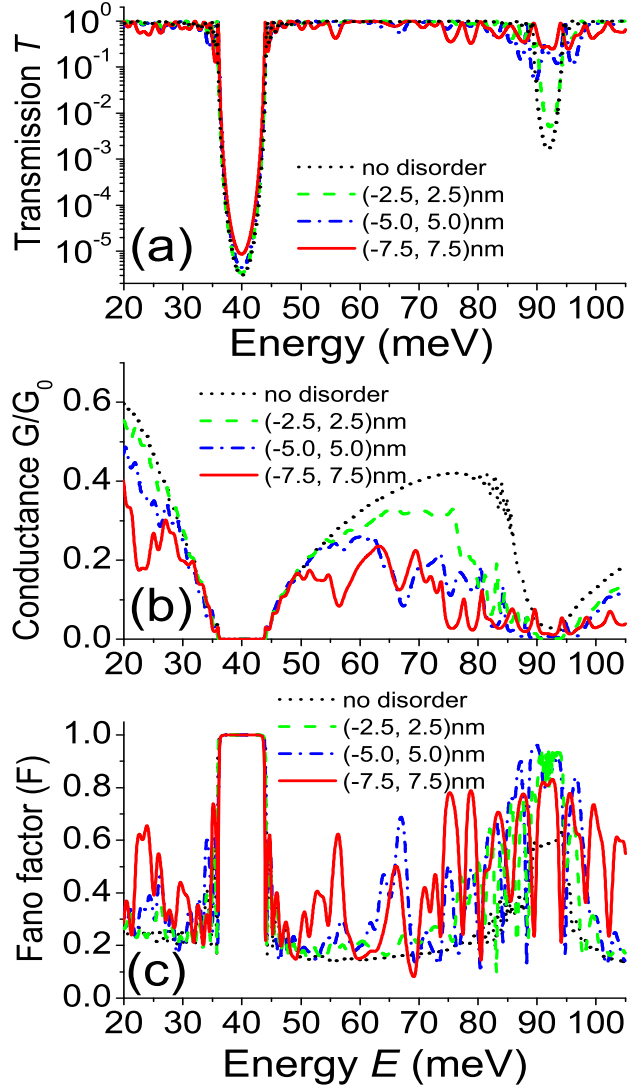}
\caption{(Color online). The effect of the structural disorder on (a)
electronic transimitivity $T=|t|^{2}$, (b) conductance $G/G_{0}$, and (c)
Fano factor, for the gapped graphene superlattice with $\Delta =5$meV, $%
V_{A}=80$meV and $V_{B}=0,$ and $w_{A}=w_{B}=(20+R)$nm, where $R$ is a
random number. The short-dashed lines denote for the structure without
disorder, the dashed lines for $R\in (-2.5,2.5)$nm, the dashed-dot lines for 
$R\in (-5,5)$nm, and the solid lines for $R\in (-7.5,7.5)$nm.}
\label{fig:FIG6}
\end{figure}

\end{document}